\documentclass{article}

\usepackage{icrc2011}

\usepackage{amssymb}
\usepackage[square,numbers]{natbib}
\usepackage{amsmath}
\usepackage{wasysym}
\usepackage{multirow}

\title{VERITAS Observations of the BL Lac Object PG 1553+113 Between May 2010 and May 2011}

\newcommand{\etal}{\MakeLowercase{\textit{et al. }}} % "et al."
\shorttitle{M. Orr \etal VERITAS Observations of PG 1553+113}

\authors{M. Orr$^{1}$ for the VERITAS Collaboration$^{2}$}
\afiliations{$^1$Department of Physics and Astronomy, Iowa State University, Ames, IA 50011\\ $^2$ see J. Holder et al. (these proceedings) or http://veritas.sao.arizona.edu/conferences/authors?icrc2011}
\email{morr@iastate.edu}

\abstract{VERITAS, an array of imaging atmospheric-Cherenkov telescopes, studies blazars in the energy range between ~100 GeV and ~30 TeV.  With its excellent sensitivity at these energies, and ever-deepening source exposures, VERITAS is in a position to probe distant blazars for clear absorption signatures in their very-high-energy gamma-ray spectra due to interactions with the extragalactic background light (EBL).  We discuss results from recent VERITAS observations of PG 1553+113 ($z \negthinspace > \negthinspace 0.4$) which have resulted in the most significant very-high-energy detection ever obtained for this source. The most recent VERITAS spectral measurements are used to place an upper limit on the source redshift of $z \negthinspace < \negthinspace 0.5$ at the 95\% confidence level. Also discussed are the prospects of using these observations, along with those of other hard- spectrum blazars, to place constraints on the EBL.} 


\begin{document}
\maketitle

%% main text
\section{Introduction}
\label{sec:Intro}
PG 1553+113 is a high-frequency peaked BL Lac  object (HBL) \citep{Falomo:1994,Giommi:1995,Beckmann:2002} discovered by Green et al. (1986) \citep{Green:1986}.  Evidence of very-high-energy (VHE) gamma-ray emission from this source was first detected by HESS in 2005 \citep{Aharonian:2006}.  This was later confirmed by observations made with the MAGIC telescope in 2005 and 2006 \citep{Albert:2007}.  Due to its featureless spectrum, the redshift of PG 1553+113 remains uncertain.  However, constraints on the redshift have been continually narrowing with improved optical measurements and limits from VHE observations  (e.g., \citep{Sbarufatti:2006,Treves:2007,Aharonian:2006,Mazin:2007}).  

Recent measurements using the \textit{Hubble Space Telescope/Cosmic Origins Spectrograph} place a lower limit on the redshift of PG 1553+113 of $z \negthinspace > \negthinspace 0.4$ \citep{Danforth:2010}.  Statistical arguments are also presented placing a $1\sigma$ upper limit of $z \negthinspace \leq \negthinspace 0.58$.

%Recent measurements using the \textit{Hubble Space Telescope/Cosmic Origins Spectrograph} have produced the tightest constraints yet on the redshift of PG 1553+113 \citep{Danforth:2010}. The authors utilize spectral absorption features, due to the interstellar and intergalactic media, over the wavelength range $1135 \, \mathrm{\AA} \negthinspace < \negthinspace \lambda \negthinspace < \negthinspace 1795 \, \mathrm{\AA}$, to place a lower limit on the source's redshift.  Based on a Ly$\alpha$+O VI absorber, the authors find a lower limit of $z \negthinspace > \negthinspace 0.395$, with a somewhat larger lower limit  of $z \negthinspace > \negthinspace 0.433$ being found from a single Ly$\alpha$ line detection. Statistical arguments are also presented placing a $1\sigma$ upper limit of $z \negthinspace \leq \negthinspace 0.58$.

%The current COS data are sensitive to Ly$\alpha$ absorbers with redshifts $z \negthinspace < \negthinspace 0.47$.  However, \citep{Danforth:2010} present statistical arguments for a $1\sigma$ upper limit of $z \negthinspace \leq \negthinspace 0.58$ on the redshift based on the lack of detection of Ly$\beta$ lines at redshifts $z \negthinspace > \negthinspace 0.4$.  If one considers these arguments to be valid, the most current constraints on PG 1553+113 place its redshift in the range of $0.43 \negthinspace < \negthinspace  z \negthinspace \lesssim \negthinspace 0.58$.

These limits suggest that PG 1553+113 is one of the most distant sources ever detected at VHE, and perhaps the most distant source of all.  This makes it an incredibly interesting source for studying the extragalactic background light (EBL).  As VHE gamma rays traverse the distance between their point of origin and the Earth, they can interact with the diffuse infrared radiation field known as the EBL \citep{Gould:1967,Stecker:1992}.  The optical/infrared EBL consists of the progressive emission of galaxies and active galactic nuclei (AGN), and the absorption and re-radiation of this emission by dust (see \citep{Hauser:2001,Kashlinsky:2005} for a review).  The level of gamma-ray absorption due to the EBL increases with both gamma-ray energy and source distance.  Consequently, observed VHE spectra are increasingly softened, with respect to their intrinsic emission, as the source becomes more distant.  This effectively produces a gamma-ray horizon, beyond which gamma rays of a particular energy are fully absorbed.  % This is sometimes represented using the so-called Fazio-Stecker relation \citep{Fazio:1970,Kneiske:2004}.  

%For a source as distant as PG 1553+113, even a ``low" EBL intensity can result in the substantial absorption of gamma-rays from the source \citep{Kneiske:2010}.  The level of absorption at larger redshifts is also increasingly dependent on the assumed evolution of the EBL (see, e.g., \citep{Primack:2001,Franceschini:2008}).  This complicates any attempt to remove the effects of EBL absorption from the spectrum and obtain information regarding the intrinsic emission.  However, this very same fact can allow for more constraining limits on the EBL, as compared to a more nearby source.

%Previous measurements of PG 1553+113 made by H.E.S.S. yield a VHE spectral index of $4.46 \pm 0.34$ between $225\,$GeV and $1.3\,$TeV \citep{Aharonian:2008}.  The \textit{Fermi} 1 Year Catalog reports the spectral index as $1.66 \pm 0.03$ between $100\,$ MeV and $100\,$GeV \citep{Abdo:2010:1FGL}.\footnote{The full one year \textit{Fermi} catalog can be found online at: \\ \texttt{http://fermi.gsfc.nasa.gov/ssc/data/access/lat/1yr\_catalog/}} 

%The results presented in this paper represent the most significant detection of PG 1553+113 to date.  Sections \ref{sec:VERITASObs} and \ref{sec:FermiObs} summarize the VERITAS and \textit{Fermi} observations, respectively.  The implications for the EBL are discussed in Section \ref{sec:EBL}.  Finally, a discussion of the results, along with concluding remarks, is given in Section \ref{sec:Discussion}.

\section{Observations}
\label{sec:Observations}
Very-high-energy observations of distant blazars are a valuable tool for constraining the EBL.  As such, PG 1553+113 has been an important component in the long-term blazar science program of the Very Energetic Radiation Imaging Telescope Array System (VERITAS) collaboration.  

The VERITAS observatory is an array of four 12-meter imaging atmospheric-Cherenkov telescopes (IACTs) located at the Fred Lawrence Whipple Observatory (FLWO) in southern Arizona \citep{Holder:2008}.  Each reflector comprises 350 hexagonal mirrors following the Davies-Cotton design.  The focal plane camera consists of 499 photomultiplier tubes covering a $3.5^\circ$ field of view.  Its large collection area ($\sim \negthinspace 10^5 \, \mathrm{m}^2$), in conjunction with the stereoscopic imaging of air showers, allows VERITAS to detect very-high-energy gamma rays between energies of $100\,$GeV and $30\,$TeV with an energy resolution of $\sim \negthinspace 15 \negthickspace - \negthickspace 25\%$ and an angular resolution of $\sim \negthinspace 0.1^\circ$.  A source with a flux 1\% that of the Crab Nebula can be detected by VERITAS in $\sim \negthinspace 25$ hours with a statistical significance of 5 standard deviations ($\sigma$).

PG 1553+113 was observed by VERITAS between May 2010 and May 2011 for a total of nearly 72 hours.  Excluding bad-weather runs, and some 3-telescope data not yet analyzed, the dataset presented here consists of 50 hours of live time.  All observations were performed in \textit{wobble} mode, with the source offset from the center of the field of view by $0.5^\circ$.  This allows for simultaneous background estimation and source observation.  Four wobble directions are used (North, South, East, West) and are alternated from run to run.

%\footnote{VERITAS typically operates with $\sim \negthinspace 10\%$ deadtime , yielding $\sim \negthinspace 58\,$ hours of live time for this dataset.  However, for 3 telescope runs, each group of runs must be processed according to which telescope is missing in the array.  While some 3 telescope data is included here, not all of it has yet been processed.  The current processed dataset has a total live time of 50 hours.}

\begin{figure}[t]
	\centering
	\includegraphics[width=3.in]{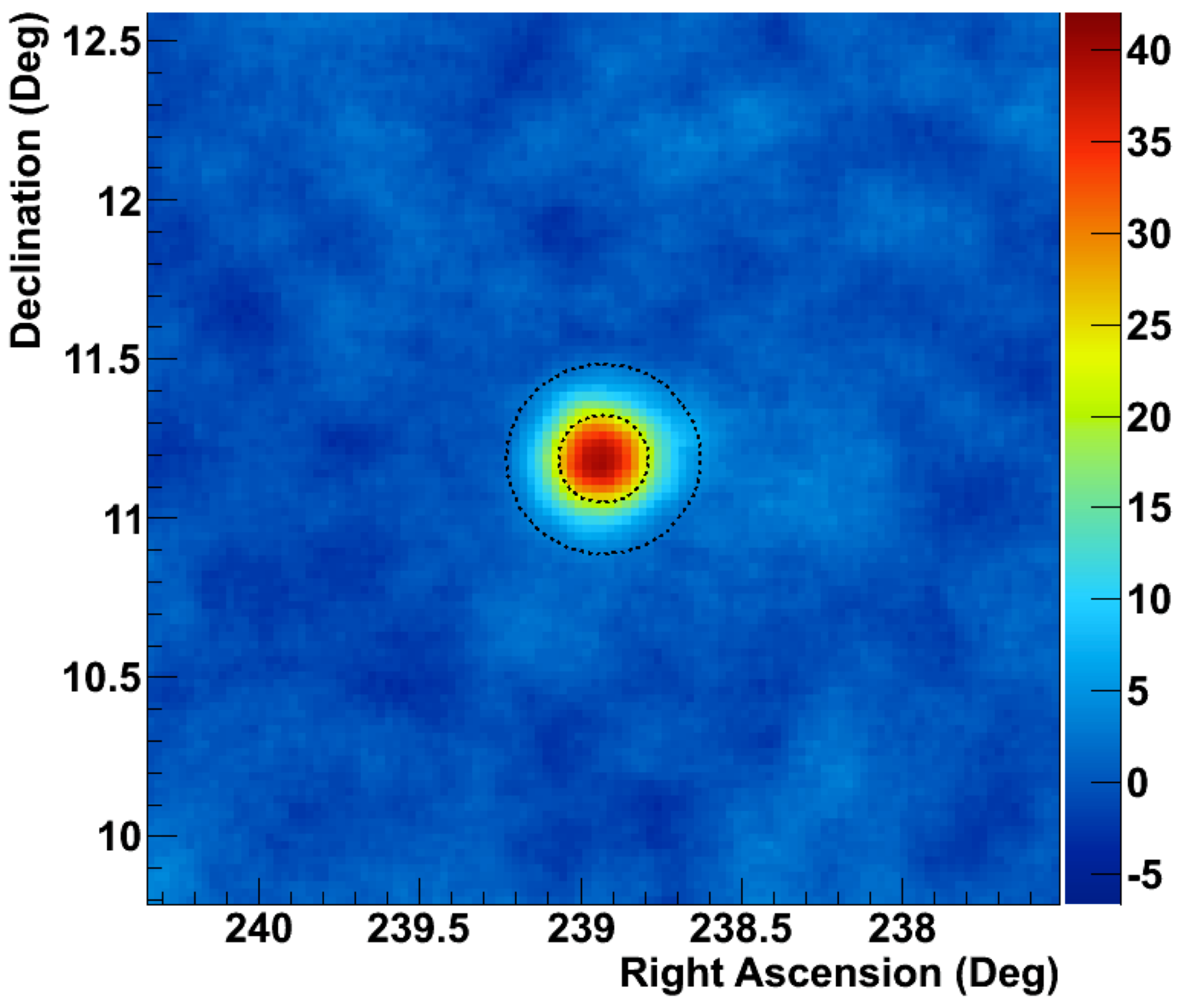}
	\caption{Significance map of PG 1553+113.  The inner dashed circle outlines the ON region used in the analysis (see Section \ref{sec:Analysis}).  The outer dashed circle indicates the region excluded from the background estimation.}
	\label{fig:skymap}
\end{figure}

\section{Analysis}
\label{sec:Analysis}
The standard VERITAS analysis chain begins with the calibration of camera gains using nightly flasher runs.  Each camera image is then cleaned, with cuts being placed on each pixel according to its signal-to-noise ratio for a given event.  The cleaned images are then characterized, using the standard Hillas parameters \citep{Hillas:1985}, by calculating the moments of the light distributions.  Cuts on these image parameters are placed to maximize the rejection of the cosmic-ray background while minimizing the loss of gamma rays.  Images from a given event passing the requisite cuts are then combined stereoscopically to reconstruct the direction of the incident gamma ray in the camera plane and the core location of the air shower in the ground plane.  The energies of reconstructed events are estimated using Monte Carlo simulations.  Separate simulations are used for data taken during the summer (May-October) and winter (November-April) months, with each set of simulations using atmospheric models appropriate for that time of year.

For the analysis of PG 1553+113, a circular region of radius $0.14^\circ$, centered around the source direction, was used to calculate the number of ON counts in the data.  The OFF counts were determined using the reflected-regions background model \citep{Berge:2007} with 7 background regions.  The detection significance was calculated following Equation 17 of \citep{Li:1983}.

\section{Results}
\label{sec:Results}
\begin{figure}[t]
	\centering
	\includegraphics[width=3.2in]{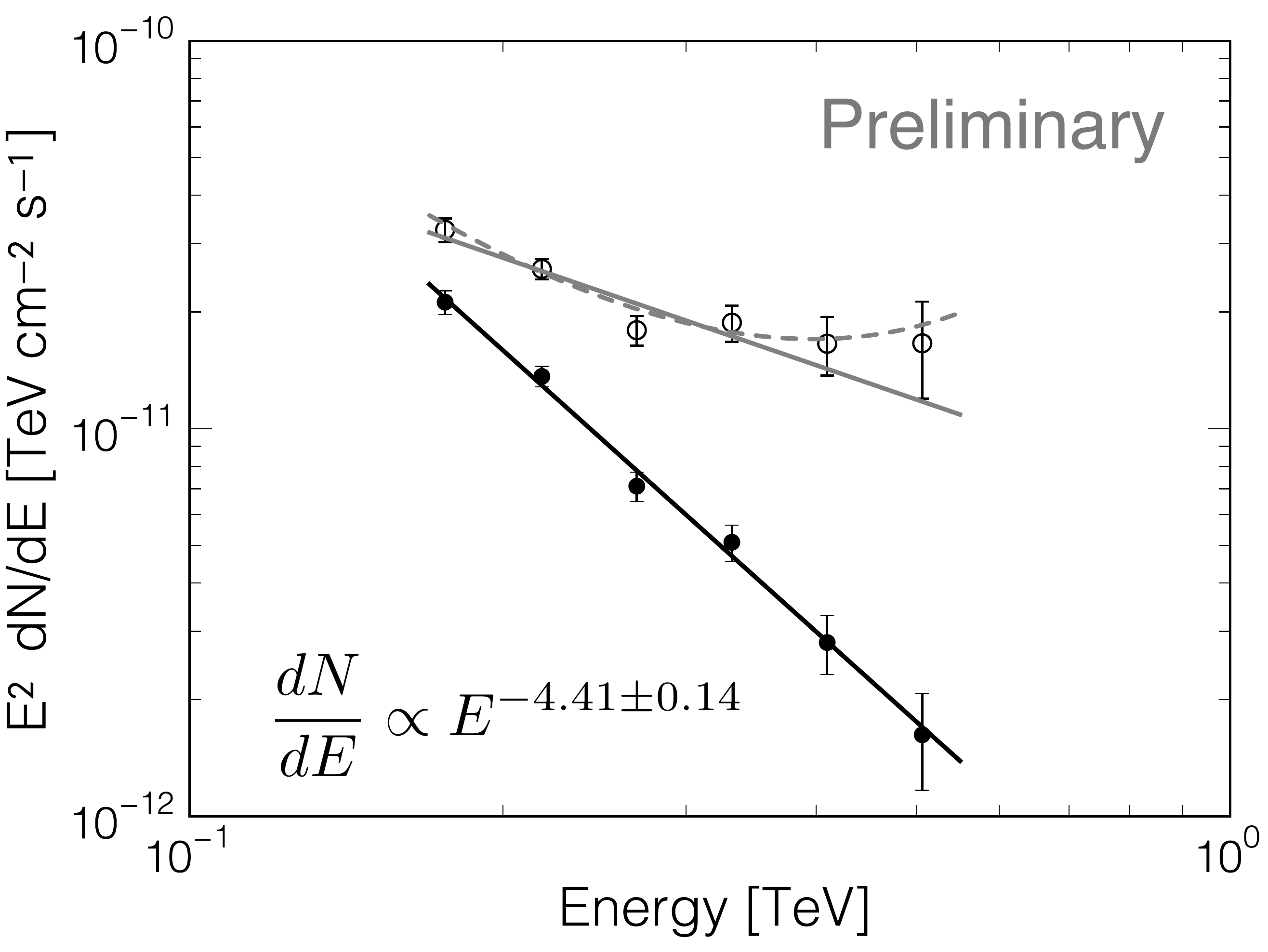}
	\caption{Observed spectrum of PG 1553+113 (solid black points and line).  The intrinsic spectrum was calculated assuming a redshift of $z=0.5$ and the low star formation rate EBL model of \citep{Kneiske:2004} is given by the open data points (see Section \ref{sec:PG1553+113Redshift}).  A simple power-law fit, and power-law fit with an exponential rise, are shown by the grey solid and dashed curves, respectively.  The flux on the y-axis is multiplied by $E^2$, where $E$ is the energy in TeV.}
	\label{fig:spectrum}
\end{figure}

The VERITAS observations of PG 1553+113 between May 2010 and May 2011 resulted in a detection significance of $39\sigma$, the most significant VHE detection ever obtained for this source. The significance map of these observations is shown in Figure \ref{fig:skymap}. The time-averaged energy spectrum for the full observation period is shown in Figure \ref{fig:spectrum}.  The spectrum spans energies from $175\,$GeV to $500\,$GeV, with all flux points having a significance greater than $3\sigma$.  A power-law fit to the spectrum yields
\begin{eqnarray}
	\label{eqn:spectrum}
	\frac{dN}{dE} = (6.66 \pm 0.34) \times 10^{-11} \left( \frac{E}{0.3\, \mathrm{TeV}} \right) ^{-4.41 \pm 0.14} \\ 
	\, \mathrm{cm}^{-2} \, \mathrm{s}^{-1} \, \mathrm{TeV}^{-1}, \nonumber
\end{eqnarray} 
with a $\chi^2/\nu$ of 2.603/4, corresponding to a fit probability of 63\%.  The time-averaged integral flux above $200\,$GeV is 
\begin{equation}
	\label{eqn:integralFlux}
	\Phi(>200 \, \mathrm{GeV}) = (2.34 \pm 0.12) \times 10^{-12} \, \mathrm{cm}^{-2} \, \mathrm{s}^{-1},
\end{equation}
where the best-fit value for the spectral index has been assumed.  This corresponds to $\sim \negthinspace 10\%$ of the Crab Nebula flux. The errors in Equations \ref{eqn:spectrum} and  \ref{eqn:integralFlux}, and all errors given below, are statistical errors only.  Table \ref{tab:ObservationSummary} provides a summary of the observations and results. The temporal analysis for the lightcurve of PG 1553+113 is currently under way.

\begin{table*}
	\caption{Summary of VERITAS observations from 2010 and 2011.}
	\footnotesize	
	\centering
	% \vspace{10pt}	
	\begin{tabular}{c c c c c c}
		\hline\hline 
		& Live Time & Significance & $\Gamma$ & $\Phi(>200 \, \mathrm{GeV})$ & \% Crab  \\
		& $\left[ \mathrm{hours} \right]$ & $\left[ \sigma \right]$ & ($dN/dE \propto E^{-\Gamma}$) & $\left[ 10^{-11} \, \mathrm{cm}^{-2} \, \mathrm{s}^{-1} \right]$ & Nebula Flux \\
		\hline
		2010 & 27 & 29 & $4.50 \pm 0.21$ & $2.43 \pm 0.14$ & 10 \\
		2011 & 23 & 26 & $4.23 \pm 0.20$ & $2.20 \pm 0.16$ & 9 \\
		\hline\hline
	\end{tabular}
	\label{tab:ObservationSummary}
\end{table*}

\section{Constraining the Redshift of PG 1553+113}
\label{sec:PG1553+113Redshift}
As previously mentioned, the redshift of PG 1553+113 is uncertain, with a firm lower limit of $z \negthinspace > \negthinspace 0.4$ placed by \citep{Danforth:2010}.  An upper limit on the redshift can be obtained using the observed VHE spectrum and correcting for attenuation due to the EBL.  There are two basic approaches to this technique.  The first is simply to calculate the de-absorbed spectrum and place a cut on the hardest (minimum) ``acceptable" value for the intrinsic VHE spectral index, thereby constraining the maximum redshift.  This value can motivated by theoretical arguments (e.g., \citep{Aharonian:2006:Nature}) or from observations of the source at lower energies where EBL attenuation is negligible (e.g., \citep{Georganopoulos:2010}).  The second approach is to calculate the de-absorbed spectrum and exclude redshifts which produce unphysical features (based on standard blazar emission models), such as an exponential rise, in the spectrum (e.g., \citep{Dwek:2005}).

With the second approach, one must have a way of determining if, for example, an apparent exponential rise is statistically significant.  This can be done using the likelihood-ratio test (see, e.g., \citep{Li:1983}).  Assuming a power-law fit to the intrinsic spectrum is the \textit{null hypothesis}, and a power law with an exponential rise (hereafter exponential power law) is the second hypothesis, the likelihood ratio can be formed as $\lambda = -2 \ln(\cal{L_\mathrm{null}} / \cal{L_\mathrm{exp}})$, where $\cal{L_\mathrm{null}}$ is the likelihood of the power-law fit and $\cal{L_\mathrm{exp}}$ is the likelihood of the exponential power-law fit.  The likelihood ratio $\lambda$ will be distributed like $\chi^2$ with $\nu_\mathrm{null} - \nu_\mathrm{exp} = 1$ degree of freedom.  The null hypothesis (power-law fit) can then be excluded, and the second hypothesis (exponential power-law fit) therefore accepted, at some confidence level (C.L.) based on the value of the likelihood ratio.

An EBL spectral energy distribution (SED) must be assumed in order for the intrinsic source spectrum to be calculated.  To obtain a true upper limit on the redshift, an SED corresponding to the lowest allowed level of EBL should be chosen, hence yielding the minimum loss of VHE gamma rays due to EBL absorption. This will result in softer calculated intrinsic spectra and weaker absorption features, thereby allowing for larger source redshifts.

Both of the aforementioned techniques were applied to PG 1553+133 by \citep{Mazin:2007} using combined MAGIC and HESS spectra. The authors derived 95\% C.L.  upper limits of $z \negthinspace < \negthinspace 0.69$ and $z \negthinspace < \negthinspace 0.42$ using the first and second approaches, respectively, and the so-called low-star-formation-rate (Low-SFR) EBL model of \citep{Kneiske:2004}.  In this work we follow the second approach.  Using the spectrum obtained from the 2010-2011 VERITAS observations, and the same EBL SED, we obtain an upper limit on the redshift of PG 1553+113 of $z \negthinspace < \negthinspace 0.5$ at the 95\% C.L.  We have followed the approach of \citep{Raue:2008} to account for the effects of EBL evolution.  The intrinsic spectrum calculated for a redhift of $z=0.5$ is shown in Figure \ref{fig:spectrum}, together with the power-law fit and exponential power-law fit.

The redshift upper limit obtained here is in good agreement with the result of \citep{Mazin:2007}.\footnote{It is important to note that constraining source redshifts using the EBL is largely dependent on the EBL SED and evolutionary scenario used.  Therefore, it is argued here that if two independent analyses, using spectra measured with different instruments, yield redshift upper limits within $\Delta z = 0.08$ of one another, this constitutes good agreement.}  The VERITAS result can be improved by expanding the observed spectrum to higher and lower energies, making the analysis more sensitive to features introduced into the intrinsic spectrum at large redshifts for a particular EBL SED.  

\section{Constraining the EBL Using PG 1553+113 and Other VERITAS Blazars}
As the redshift of PG 1553+113 becomes increasingly constrained, it will also become a valuable source for constraining the intensity of the EBL and the shape of its SED.  With a known source redshift, the same analysis described in Section \ref{sec:PG1553+113Redshift} can be applied, leaving the redshift fixed and varying the EBL intensity instead.  In this case, features in the intrinsic spectrum will result from an increase in EBL intensity rather than an increase in redshift.  

\begin{table*}[t]
	\caption{Blazars in the VERITAS catalog well suited for constraining the EBL.   Listed are (left to right) the source name, redshift, luminosity distance, LAT spectral index, VERITAS spectral index, and the difference in spectral indices divided by the luminosity distance.  All \textit{Fermi}-LAT spectral indices are taken from the 1FGL catalog (\texttt{\small{http://fermi.gsfc.nasa.gov/ssc/data/access/lat/1yr\_catalog/}}).}
	\footnotesize	
	\centering
	\vspace{10pt}	
	\begin{tabular}{c c c c c c}
		\hline\hline 
		Source Name & Redshift & $D_L$ & $\Gamma_\mathrm{LAT}$ & $\Gamma_\mathrm{VTS}$ &  $\alpha = \Delta\Gamma/D_L$ \\
		& & $\left[ \mathrm{Mpc} \right]$  & & & $\left[ \times 10^{-3} \, \mathrm{Mpc}^{-1} \right]$ \\
		\hline
		RGB J0710+591 & 0.124 & 488  & $1.28 \pm 0.21$ & $2.69 \pm 0.26$ & 2.9 \\
		1ES 0229+200 & 0.139 & 542 & - & ~ $2.44 \pm 0.11^\dagger$ & ~ $1.9^\ddagger$ \\
		1ES 1218+304 & 0.182 & 689 & $1.70 \pm 0.08$ & $3.07 \pm 0.09$ & 2.0 \\
		RBS 0413 & 0.190 & 716 & $1.47 \pm 0.18$ & ~ $3.24 \pm 0.84^\dagger$ & 2.5 \\
		1ES 0414+009 & 0.287 & 1017 & $1.94 \pm 0.22$ & ~ $3.05 \pm 0.41^\dagger$ & 1.2 \\
		PG 1553+113 & 0.4--0.5 & 1323--1560 & $1.66 \pm 0.03$ & $4.41 \pm 0.14$ & 1.8--2.1 \\
		\hline\hline
		\multicolumn{6}{l}{$^\dagger$ Preliminary.} \\
		% \multicolumn{6}{l}{$^*$ Using $z=0.45$.}  \\
		\multicolumn{6}{l}{$^\ddagger$ Assuming $\Gamma_\mathrm{LAT} = 1.5$.}
	\end{tabular}
	\label{tab:BlazarList}
\end{table*}

As briefly mentioned in Section \ref{sec:PG1553+113Redshift}, one approach to constraining the EBL is to use blazar spectra measured at MeV to GeV energies, where EBL attenuation is negligible, as a lower limit to the hardness of the intrinsic TeV spectrum.  This method yields reliable results as long as the spectrum does not exhibit an additional TeV spectral component. The \textit{Fermi} Large Area Telescope (LAT), operating in the energy range of $\sim \negthinspace 100\,$MeV to $300\,$GeV, is ideal for establishing a proxy for intrinsic TeV spectral limits.  The degree to which a particular source can constrain the EBL can be quantified using the ratio $\alpha = \Delta\Gamma/D_L$ \citep{Georganopoulos:2010}, where $\Delta\Gamma$ is the observed VERITAS spectral index ($\Gamma_\mathrm{VTS}$) minus the LAT measured spectral index ($\Gamma_\mathrm{LAT}$) and $D_L$ is the luminosity distance of the source.  A smaller value of $\alpha$ indicates the source will more tightly constrain the EBL.

A number of blazars in the VERITAS catalog are well suited for constraining the EBL, based on their small values of $\alpha$.  These sources are shown in Table \ref{tab:BlazarList}.  It can be seen that all of the sources listed have values of $\alpha \negthinspace < \negthinspace 3 \times 10^{-3} \, \mathrm{Mpc}^{-1}$.  For comparison, the nearby blazar 1ES 2344+514 has $\alpha = (2.95-1.57)/183 \,\mathrm{Mpc} = 7.5 \times 10^{-3} \, \mathrm{Mpc}^{-1}$, indicating that it is not an ideal source for constraining the EBL using this technique.

By combining the individual limits on the EBL obtained using each source from Table \ref{tab:BlazarList}, the constraints can be further improved.  Many of these sources (PG 1553+113 included) are components of the VERITAS blazar long-term observing plan and will have improved spectral measurement over the years to come.  This will continue to tighten the region of parameter space allowed for the intensity and shape of the EBL.

\section{Discussion \& Conclusions}
\label{sec:Discussion}
The VERITAS observations of PG 1553+113 between May 2010 and May 2011 have resulted in the most significant VHE detection of this source to date.  A time-averaged observed spectral index of $\Gamma = 4.41 \pm 0.14$ is obtained for the full data set.  The time-averaged integral fluxes above $200\,$GeV for the 2010 and 2011 time periods show no evidence for annual variability.  A shorter timescale ($\sim\,$daily) variability analysis is currently under way. 

We derive an upper limit for the redshift of PG 1553+113 of $z \negthinspace < \negthinspace 0.5$, at the 95\% C.L., based on the assumption that the intrinsic spectrum does not  rise exponentially with increasing energy above $\sim \negthinspace 300\,$GeV. Combined with the lower limit of \cite{Danforth:2010}, this places the redshift in the range $ 0.4 \negthinspace < \negthinspace z \negthinspace < \negthinspace 0.5$.  As the limits on the redshift of PG 1553+113 tighten, this object will become an effective tool for constraining the EBL.

\section*{Acknowledgements}
This research is supported by grants from the US Department of Energy, the US National Science Foundation, and the Smithsonian Institution, by NSERC in Canada, by Science Foundation Ireland, and by STFC in the UK. We acknowledge the excellent work of the technical support staff at the FLWO and at the collaborating institutions in the construction and operation of the instrument.

%\bibliographystyle{apj}
%\bibliography{apj-jour,bibliography}

\clearpage

\end{document}